\documentclass[aps,prl,reprint,superscriptaddress,onecolumn,showpacs,notitlepage]{revtex4-1}
\setcitestyle{numbers,square}

\usepackage[T1]{fontenc}
\usepackage[utf8]{inputenc}
\usepackage{textcomp}
\usepackage{amsmath}
\usepackage{mathtools,amssymb,booktabs}

\usepackage{gensymb}
\usepackage{multirow}

\usepackage{hyperref}
\hypersetup{
  colorlinks=true,
  linkcolor	=black,
  citecolor	=blue,
  urlcolor	=blue
}

\newcommand{\muB}{\mu_\mathrm{B}}

\newcommand{\Tc}{T_\mathrm{c}}
\newcommand{\quotemarks}[1]{``#1''}
\newcommand{\hakasulut}[1]{\left[#1\right]}
\newcommand{\Egap}{E_\text{gap}}

\begin{document}

\title{Magnetic properties of polyacetylene: Exploring electronic correlation effects through first-principles modeling}

\author{Johannes Nokelainen}
\email{johannes.nokelainen@lut.fi}
\affiliation{Department of Physics, School of Engineering Science, LUT University, 53850 Lappeenranta, Finland}
\affiliation{Department of Mechanical Engineering, School of Energy Systems, LUT University, 53850 Lappeenranta, Finland}
\affiliation{Department of Physics, Northeastern University, Boston, Massachusetts 02115, USA}
\affiliation{Quantum Materials and Sensing Institute, Northeastern University, Burlington, Massachusetts 01803, USA}

\author{Bernardo Barbiellini}
\affiliation{Department of Physics, School of Engineering Science, LUT University, 53850 Lappeenranta, Finland}
\affiliation{Department of Physics, Northeastern University, Boston, Massachusetts 02115, USA}
\affiliation{Quantum Materials and Sensing Institute, Northeastern University, Burlington, Massachusetts 01803, USA}

\author{Arun Bansil}
\email{ar.bansil@northeastern.edu}
\affiliation{Department of Physics, Northeastern University, Boston, Massachusetts 02115, USA}
\affiliation{Quantum Materials and Sensing Institute, Northeastern University, Burlington, Massachusetts 01803, USA}

\date{\today}

\begin{abstract}
  Polyacetylene,
  a simple yet fascinating polymer,
  has been of great interest for its unique electronic properties.
  However,
  the role of electronic correlation effects in polyacetylene still has not been explored fully on an \emph{ab initio} basis.
  Using density functional theory (DFT) and a range of exchange-correlation functionals---including GGA, meta-GGA, and hybrid functionals---we demonstrate that correlation effects can possibly stabilize a magnetic state as a competing order on the $\pi$-conjugated carbon $p$ orbitals.
  Our study highlights the complexity of physics of polyacetylene and suggests similarities with the physics of the cuprates.
\end{abstract}

\maketitle

\section{Introduction}

Interest in conjugated polymers is driven by their unique properties as low-dimensional
solids, where electron-electron and electron-lattice interactions play a critical role in shaping
the electronic structure, and present a challenge for understanding the complex underlying
physics~\cite{1992_Kiess_polymer_book, 2005_Barford_polymer_book}.
Conjugated polymers are classified as quasi-one-dimensional systems, where electronic
properties differ substantially from those of the traditional inorganic semiconductors such as
silicon. Remarkably, both silicon~\cite{Barbiellini_1985} and conjugated
polymers~\cite{beckerle_2000} have been successfully utilized in particle detection,
showcasing the diverse applications of these materials with distinct electronic characteristics.

\begin{figure}[htpb]
 \centering
 \includegraphics[width=.4\linewidth]{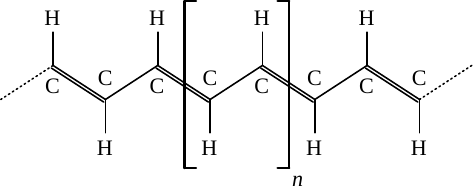}
 \caption{
  Structure and the computational unit cell of the infinite \emph{trans}-polyacetylene chain $\hakasulut{\text{C}_2\text{H}_2}_n$
  (other isomers of polyacetylene were not considered in this study).
  In the dimerized CDW state the polyacetylene chain consists of alternating expanded double bonds and contracted single bonds.
}
\label{fig:structure}
\end{figure}

Polyacetylene (Fig.~\ref{fig:structure}) is one of the simplest polymers.
It was the first organic compound to exhibit a remarkable increase in conductivity with electron-doping,
turning from an insulating to a metallic state~\cite{1977_Chiang_polyacetylene_conductivity}.
This discovery opened the field of organic microelectronics and was recognized by the Nobel Prize in Chemistry in 2000~\cite{2001_Shirakawa_polyacetylene_nobel_lecture}.
It features a semiconducting electronic structure with a band gap of about 2.4\,eV in its $\pi$-bonded carbon $p$ bands,
as recently verified directly by ARPES experiments~\cite{2019_Wang_polyacetylene_EXP}.
The Su--Schrieffer--Heeger (SSH) model~\cite{1979_SSH,
1979_Rice_solitons_polyacetylene,
2000_Ramasesha_review_conjugated_polymers_density_matrix_renormalization}
has provided an explanation for the band gap and the doping-induced increase in conductivity:
dimerization of the polyacetylene chain due to Peierls distortion leads to the formation of a charge density wave (CDW),
which opens the band gap and drives mid-gap soliton excitations that account for the conductivity.
This picture has been supported by first principles calculations~\cite{1998_Champagne_polyacetylene_HF_DFT,
1998_Choi_polyacetylene_correlation_BLA,
2012_Lacivita_polyacetylene_DFT_polarizability,
2014_Chattopadhyaya_polymer_DFT_band_gap_correlation,
2015_Barborini_polyacetylene_QMC_RVB_hybrid_functionals}
and the experimental work showing the presence of solitons~\cite{1988_Heeger_review_soliton_polyacetylene,
1994_Soos_polymer_fluorescene_structure}.
However,
direct experimental evidence for bond alternation is limited and has been disputed by Hudson~\cite{2013_Hudson_bond_alternation,2018_Hudson_polyacetylene_review}.

It is well-known that if the value of the on-site Coulomb correlation energy on carbon sites is high,
the band gap in polyacetylene could also be explained by the formation of a spin density wave (SDW) state~\cite{1969_Harris_polyacetylene_charge_spin_density_waves,
1981_Subbaswamy_polyacetylene_correlation,
1982_Fukutome_polyacetylene_charge_spin_density_waves}.
This picture is familiar from the cuprate high-$\Tc$ superconductors,
where electronic correlations give rise to a Mott insulator state in which the Cu magnetic moments are arranged in an antiferromagnetic (AFM) network~\cite{2006_Lee_review_high-Tc_physics}.
However,
some theoretical studies have concluded that the role of correlation is merely to affect the bond-length alternation of the dimerized carbon backbone and that the CDW phase is preferred over the SDW state for any value of correlation strength~\cite{1981_Horsch_polyacetylene_correlation_BLA,
1985_Baeriswyl_polyacetylene_correlation}.
Furthermore,
the magnetic properties have not been considered in prior first-principles studies~\cite{1998_Champagne_polyacetylene_HF_DFT,
2002_Ma_polyacetylene_time_dependent_DFT,
2015_Barborini_polyacetylene_QMC_RVB_hybrid_functionals,
2023_Furuya_polyacetylene_DFT_vibrations}
and the existing experimental studies often focus on magnetic interactions between the magnetic impurities~\cite{1990_Winter_polyacetylene_free_radicals_magnetism,1991_Cosmo_polyacetylene_free_radicals_magnetism} rather than intrinsic magnetism.

On this account,
despite extensive studies,
there still seem to be gaps in understanding of role of electronic correlations on the structural and magnetic properties of polyacetylene.
It is with this motivation that here we use the density-functional theory (DFT) to explore the role of electronic correlations in polyacetylene on an \emph{ab initio} basis.
For this purpose,
we employ various exchange-correlation (XC) functionals that provide different approximative descriptions for XC effects.
These functionals include some of the most popular and well-established functionals that represent various rungs of the so-called \quotemarks{Jacob's ladder of XC functionals},
where climbing up the ladder provides systematic increase in the complexity of the functionals~\cite{2001_Perdew_DFT_Jacobs_ladder}.
In the rung of the generalized gradient approximation (GGA) we use the PBE functional~\cite{1996_PBE_functional} as well as $\text{PBE}+U$,
where electronic correlations are included via the addition of an external Hubbard $U$ parameter that controls a simple penalty potential for the atomic orbitals.
In the meta-GGA level we use the SCAN functional~\cite{2015_Sun_SCAN},
which has been shown to provide a systematically improved parameter-free treatment of electronic correlations in diverse materials~\cite{2018_Chris_La2CuO4_SCAN,
2018_Isaacs_SCAN_performance_solids,
2020_Nokelainen_bisco,
2020_Yubo_SCAN_stripe_YBCO,
2022_Kanun_functional_comparisons,
2024_Nokelainen_ybco_pressure,
2024_Ruiqi_nickelate_nemacity},
and the revised r$^2$SCAN functional~\cite{2020_Furness_r2SCAN_functional}.
In the rung of hybrid functionals we have chosen the popular HSE06~\cite{2003_HSE_functional} and B3LYP~\cite{1993_Becke_B3LYP_part1,1994_Stephens_B3LYP_part2} functionals.

\section{Methodology}

The DFT simulations were carried out using the Vienna \emph{ab initio} simulation package (VASP)~\cite{1996_Kresse_VASP_PRB} based on the projector-augmented-wave formalism~\cite{Blochl1994_PAW}.
A plane-wave basis set with a kinetic energy cutoff of 550\,eV was used and the electronic energy minimization was performed with a tolerance of 10$^{-6}$\,eV.
A Gaussian smearing width of 0.2\,eV (FWHM) was applied to the electronic states.
The one-dimensional Brillouin zone of an infinite polyacetylene chain was sampled with a $\Gamma$-centered $\mathbf{k}$-grid consisting of
as many as 100 points in order to make sure that the grid density is sufficient also for the B3LYP functional.
We separated the polyacetylene chains by a vacuum of at least 10\,Å and relaxed the structures for each tested functional until the atomic forces were below
0.01\,eV/Å.
To ensure that the system was not sitting on top of an unstable maximum point in the potential energy landscape,
we used reduced-symmetry geometry (CDW phase structure file) and reduced-symmetry electronic state (SDW phase charge density and wave function files) as the starting point for all the relaxations.
In the $\text{PBE}+U$ calculations the Hubbard parameter $U$ was applied on the $p$ orbitals with the rotationally invariant approach of Ref.~\cite{1998_Dudarev_DFT+U_LDAUTYPE2}.

\section{Results and discussion}

Our benchmarking tests of various XC fuctionals on polyacetylene revealed surprising results that are presented and compared with existing literature in Table~\ref{tab:table}.
Firstly,
the bond alternation was found to be zero for most of the XC functionals studied.
Previously the dimerized CDW geometry has been found for PBE and HSE06 and even for the local density approximation (LDA)~\cite{1989_Ashkenazi_polyacetylene_Peierls,
2012_Lacivita_polyacetylene_DFT_polarizability,
2014_Chattopadhyaya_polymer_DFT_band_gap_correlation}
but in our case the relaxation converged to the CDW state only for the B3LYP hybrid functional.
Secondly,
the emergence of a SDW state with antiferromagnetically ordered C magnetic moments was observed with $\text{PBE}+U$, SCAN, r$^2$SCAN and HSE06,
which has not been reported before in earlier \emph{ab initio} studies,
possibly due to the omission of spin degree of freedom in the computations.


It is particularly surprising that in the case of PBE neither bond alternation nor magnetic solutions were found since according to the Peierls theorem,
the ground state is expected to be a broken-symmetry phase.
Interestingly,
Ref.~\cite{2002_Sun_polyacetylene_gating_DFT} found for the PW91 GGA functional that bond alternation is not found for high $\mathbf{k}$-point densities (as in our case),
but it is found for lower densities that can be interpreted as more \quotemarks{molecular} description of the polyacetylene chain,
rather than considering it as a perfectly repeating infinite chain when a high $\mathbf{k}$-point density is used.
Therefore,
it could be that the previous reports such as Refs.~\cite{1989_Ashkenazi_polyacetylene_Peierls,
2012_Lacivita_polyacetylene_DFT_polarizability} used insufficient number of $k$-points,
leading to bond alternation for the LDA and PBE functionals.

\begin{table}[tpb]
\caption{Polyacetylene structural parameters and magnetic moments for all of the functionals studied here and also from literature.
Angle $\theta_\text{CCC}$ stands for the bond angle of the carbon backbone,
i.e.,
the C---C---C bond angle.
The angle $\theta_\text{CCH}$ is the angle between the longer C---C bond
(in the case of the CDW phase with expanded double bonds, see Fig.~\ref{fig:structure})
and the C---H bond.
In the symmetric non-dimerized case the C---C bond lengths are equal and $\theta_\text{CCH} = (360\degree - \theta_\text{CCH})/2$,
so the shorter C---C bond length and $\theta_\text{CCH}$ are not independent variables and we mark them with \quotemarks{symm.}.
The results of Ref.~\cite{2015_Barborini_polyacetylene_QMC_RVB_hybrid_functionals} refer to the values at the middle of the longest studied finite chain that has the length of 24 C atoms.
}
\newcommand{\duplicate}{---$\parallel$---}
\newcommand{\symm}{symm.}
\centering
\begin{tabular}{@{\extracolsep{4pt}}lccccccc}
\toprule
 \multirow{3}{*}{Functional} &
 \multicolumn{3}{c}{Bond lengths (Å)} &
 \multicolumn{2}{c}{Bond angles (\degree)} &
 \multirow{3}{*}{$M$ ($\muB$)} &
 \multirow{3}{*}{$\Egap$ (eV)}
 \\
 \cmidrule{2-4} \cmidrule{5-6}
 & \multicolumn{2}{c}{$l_\text{CC}$} & \multirow{2}{*}{$l_\text{CH}$} & \multirow{2}{*}{$\theta_\text{CCC}$} & \multirow{2}{*}{$\theta_\text{CCH}$} \\
 \cmidrule{2-3}
 & long & short & & & \\
 \hline
 PBE &
 1.396 & \symm & 1.096 & 124.5 & \symm & 0.000 & 0.00
 \\
 $\text{PBE} + U = 3$\,eV &
 1.415 & \symm & 1.104 & 124.3 & \symm & 0.032 & 0.20
 \\
 $\text{PBE} + U = 6$\,eV &
 1.436 & \symm & 1.113 & 124.0 & \symm & 0.107 & 0.77 
 \\
 $\text{PBE} + U = 9$\,eV &
 1.461 & \symm & 1.124 & 123.6 & \symm & 0.168 & 1.37
 \\
 SCAN &
 1.390 & \symm & 1.088 & 124.3 & \symm & 0.164 & 1.59
 \\
 r$^2$SCAN &
 1.392 & \symm & 1.089 & 124.4 & \symm & 0.122 & 1.03
 \\
 HSE06 &
 1.377 & \symm & 1.062 & 126.4 & \symm & 0.102 & 1.05
 \\
 B3LYP &
 1.491 & 1.382 & 1.062 & 131.0 & 115.7 & 0.000 & 3.38
 \\
 \midrule
 PBE~\cite{2012_Lacivita_polyacetylene_DFT_polarizability} &
 1.412 & 1.399 & & & & & 0.114
 \\
 HSE06~\cite{2014_Chattopadhyaya_polymer_DFT_band_gap_correlation} &
 1.45 & 1.36 & & & & & 2.27
 \\
 B3LYP~\cite{2012_Lacivita_polyacetylene_DFT_polarizability} &
 1.427 & 1.374 & & & & & 1.167
 \\
 B3LYP~\cite{2014_Chattopadhyaya_polymer_DFT_band_gap_correlation} &
 1.46 & 1.34 & & & & & 2.48
 \\
 B3LYP~\cite{2015_Barborini_polyacetylene_QMC_RVB_hybrid_functionals} &
 1.428 & 1.367 & & & &
 \\
 HF~\cite{1998_Champagne_polyacetylene_HF_DFT} &
 1.451 & 1.339 & 1.077 & 124.5 & 116.3 &
 \\
 HF~\cite{2012_Lacivita_polyacetylene_DFT_polarizability} &
 1.452 & 1.338 & & & & & 6.805
 \\
 HF~\cite{2015_Barborini_polyacetylene_QMC_RVB_hybrid_functionals} &
 1.455 & 1.333 & & & &
 \\
 MP2~\cite{2015_Barborini_polyacetylene_QMC_RVB_hybrid_functionals} &
 1.440 & 1.346 & & & &
 \\
 QMC~\cite{2015_Barborini_polyacetylene_QMC_RVB_hybrid_functionals} &
 1.431 & 1.366 & & & &
 \\
 \bottomrule
\end{tabular}
\label{tab:table}
\end{table}

\begin{figure}[tb]
 \centering
 \includegraphics[width=.7\linewidth]{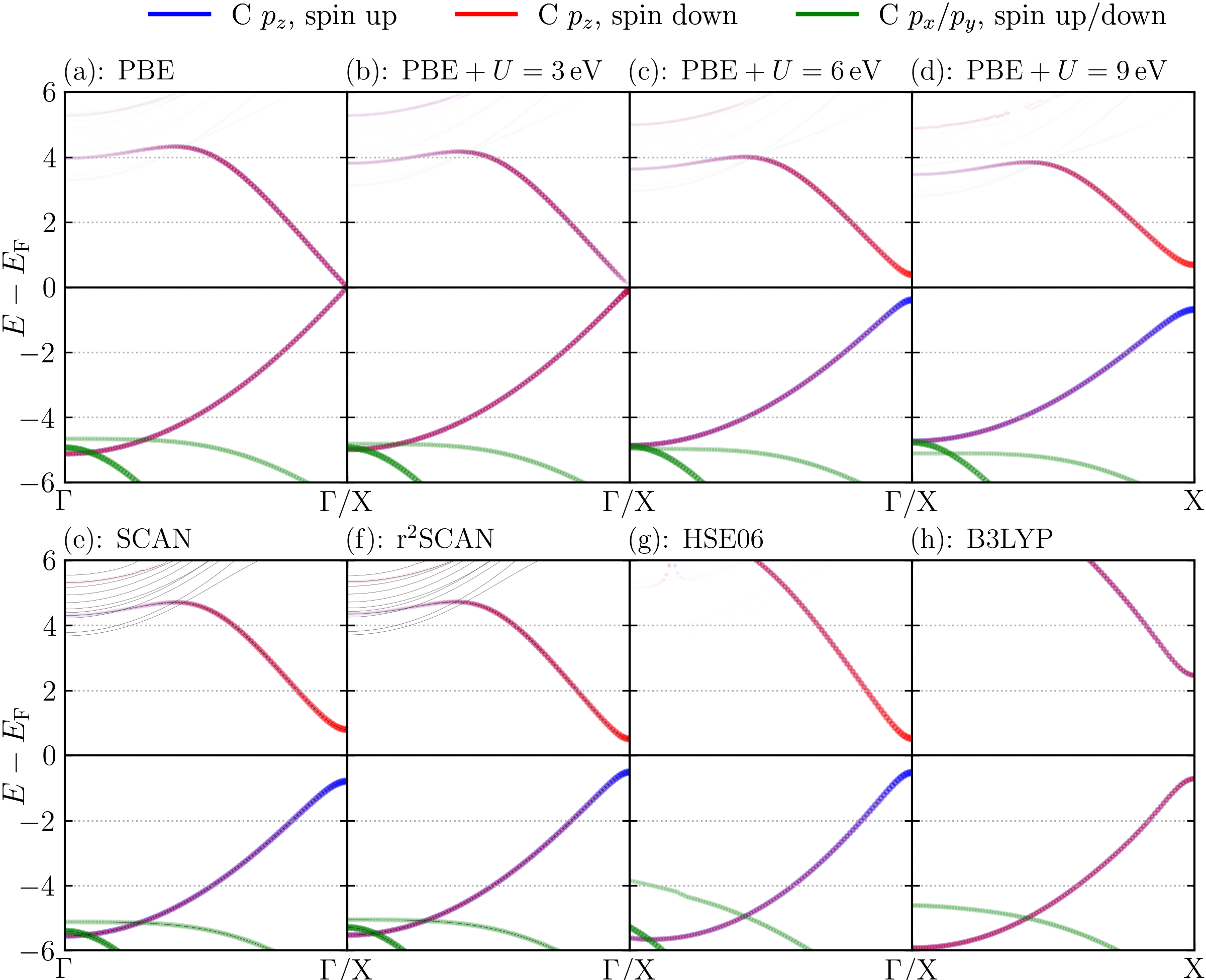}
 \caption{
  Orbital and site-decomposed band dispersions along the special symmetry line between $\Gamma=0$ and $X=0.5$ for the investigated functionals.
  The site-decomposition has been performed on an individual C site with $M \geq 0$ so that the spin-polarization of the $p_z$ bands becomes visible.
}
\label{fig:bands}
\end{figure}
\begin{figure}[tb]
 \centering
 \includegraphics[width=.7\linewidth]{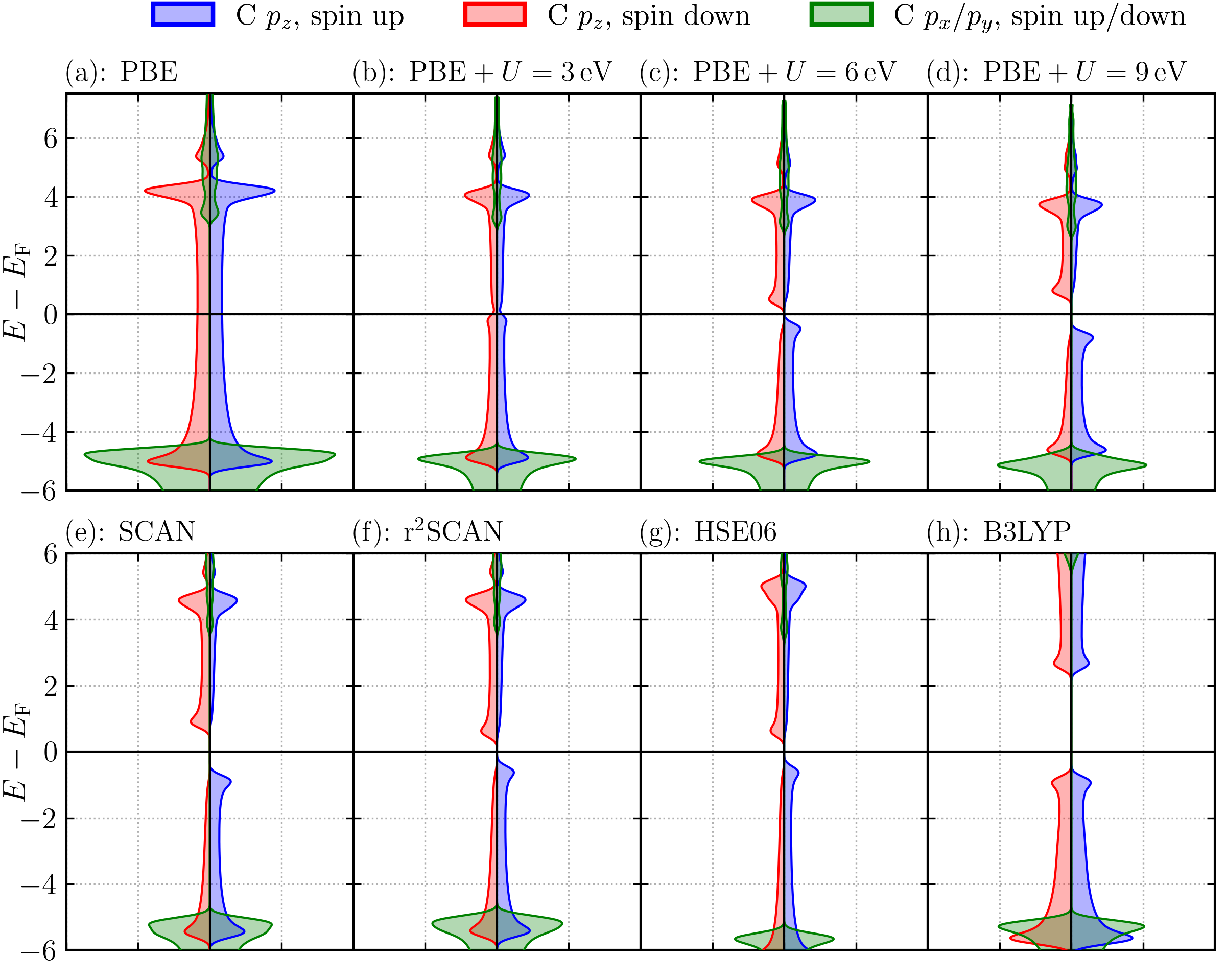}
 \caption{
  Orbital and site-decomposed PDOS for the investigated functionals.
  The site-decomposition has been performed on an individual C site with $M \geq 0$ so that the spin-polarization of the $p_z$ bands becomes visible.
}
\label{fig:dos}
\end{figure}

The band structure and partial density of states (PDOS) for each functional studied are visualized in Figs.~\ref{fig:bands} and Fig.~\ref{fig:dos},
respectively.
The spectrum is gapless for PBE since it does not feature either of the symmetry-breaking mechanisms (CDW nor SDW),
but in accordance with the Peierls theorem,
both symmetry-breaking mechanisms lead to the formation of a band gap in the $\pi$-bonding $p_z$ bands.
In the case of the SDW state solutions,
the C magnetic moments are caused by the site-projected $p_z$ bands that have spin up / down polarization for the valence/conduction bands like the Mott insulator state corresponding to the $d_{x^2-y^2}$ band in the cuprates~\cite{2020_Nokelainen_bisco}.


The $\text{PBE}+U$ method provides the most straightforward insight into the effects of correlation since it includes the correlation directly
(albeit approximately in a rotationally invariant manner~\cite{1998_Dudarev_DFT+U_LDAUTYPE2})
through the Hubbard parameter $U$,
which controls the strength of a penalty functional that promotes orbital localization.
As expected,
the band gap and the magnetic moment are proportional to the strength of the applied $U$ potential.
We obtain $\Egap = 0.20$\,eV, 0.77\,eV, 1.37\,eV and $M=0.032\,\muB$, $0.107\,\muB$, $0.167\,\muB$ for $U=3$\,eV, 6\,eV, 9\,eV.
This confirms the theoretical result that the electronic correlation is the driving force behind the formation of the SDW state like the correlated AFM Mott insulator state in the cuprates~\cite{2006_Lee_review_high-Tc_physics}.
Furthermore,
the SCAN functional,
which is well-known for being able to provide a good description of aspects of the cuprates and other strongly correlated materials~\cite{2018_Chris_La2CuO4_SCAN,
    2018_Isaacs_SCAN_performance_solids,
    2020_Nokelainen_bisco,
    2020_Yubo_SCAN_stripe_YBCO,
    2022_Kanun_functional_comparisons,
    2024_Nokelainen_ybco_pressure,
    2024_Ruiqi_nickelate_nemacity},
predicts $M=0.164\,\muB$ and $\Egap = 1.59$\,eV.
SCAN and $\text{PBE}+U=9$\,eV yield similar magnetic moments and band gaps,
which implies that the strength of the correlation contained in the SCAN functional for this system is around 9\,eV.
The revised r$^2$SCAN functional and the hybrid functional HSE06 predict smaller but similar values for $M$ ($0.122\,\muB$ and $0.102\,\muB$) and $\Egap$ ($0.792$\,eV and $0.743$\,eV).
Comparing these values with the $\text{PBE}+U=6$\,eV results indicates that these functionals also contain more than 6\,eV of correlation for this system.
These values are larger than for example in the $d$-electron systems such as cuprates,
for example a value of 4.8\,eV is estimated for La$_2$CuO$_4$~\cite{2018_Chris_La2CuO4_SCAN}.
However,
these values still are realistic.
In fact,
Baeriswyl and Maki have predicted $U=7$--9\,eV~\cite{1985_Baeriswyl_polyacetylene_correlation},
albeit their model is based on the CDW state rather than the SDW state.
B3LYP yields an exaggerated band gap of 3.38\,eV,
which is also clearly larger than the values reported in the literature~\cite{2012_Lacivita_polyacetylene_DFT_polarizability,
2014_Chattopadhyaya_polymer_DFT_band_gap_correlation}.
This could be due to the particularly dense $k$-mesh used in our study,
as we noticed that sparser $k$-mesh leads to smaller $\Egap$.


We have tabulated the prediction of structural parameters for different functionals in Table~\ref{tab:table}.
The dimerized CDW geometry given by our B3LYP simulations is in reasonable agreement with the earlier literature, which reported results on B3LYP, Hartree-Fock (HF), MP2, and Quantum Monte Carlo (QMC),
although the C---C bond lengths ($l_\text{CC}$) are longer than previously reported.
Moving on to the higher-symmetry geometries obtained with rest of the functionals,
increasing the correlation through the Hubbard $U$ parameter in $\text{PBE}+U$ significantly elongates the C---C and C---H bonds and decreases the bond angle of the carbon backbone chain ($\theta_\text{CCC}$).
On the contrary,
compared to PBE,
SCAN and r$^2$SCAN predict similar $l_\text{CC}$, slightly reduced $l_\text{CH}$ (C---H bond length) and similar $\theta_\text{CCC}$.
HSE06 contracts both C---C and C---H bonds and increases $\theta_\text{CCC}$ compared to PBE.
Therefore,
the correlation appears to have opposite effects on the structural properties when applied through the $\text{PBE}+U$ method than when applied through more advanced functionals.
However,
in the case of $\text{PBE}+U$ the expansion of the bonds could be caused by decreased covalency due to Hubbard $U$,
which is applied on all $p$ orbitals alike,
leading to localization of the $p_{x}/p_{y}$ orbitals.
This is likely to be a spurious effect since PBE has traditionally provided a good description of the carbon $sp^2$ and $sp^3$ bonding physics~\cite{2016_Lechner_carbon_DFT_benchmarking},
implying that the covalent $sp^2$ and $sp^3$ bonds do not contain any correlation physics beyond the PBE.
Correlation is thus likely to contract the bonds and increase bond angle of the carbon backbone chain,
as predicted by SCAN, r$^2$SCAN and HSE06 where the correlation has been included at a higher level.

While the CDW state predicted by the B3LYP hybrid functional is the generally accepted ground state for polyacetylene,
it is noteworthy how the AFM solutions are robustly found with a wide range of functionals belonging to various rungs of Jacob's ladder,
including the HSE06 hybrid functional.
Thus,
the SDW state is also likely to have physical significance in the sense that even if it is not the true ground state of polyacetylene,
both the SDW state and the CDW state might be possible to stabilize at different parts of the polyacetylene phase diagram as a function of doping or strains,
or that these states could belong to competing,
possibly intertwined,
phases in polyacetylene.
It could be that the CDW and SDW could form more complex fluctuating stripe phases observed in the cuprates~\cite{2020_Yubo_SCAN_stripe_YBCO} that could be stabilized in larger computational supercells.
It could also be that several phases could coexist and together contribute to the band gap,
possibly featuring time fluctuations,
reminiscent of the pseudogap phase hosted by the cuprates.
Pseudogap is an anomalous gap that exists in the cuprate normal-state energy spectrum above $\Tc$.
While the nature of this phase remains controversial,
it is widely believed to be composed of intertwined AFM and CDW/SDW orders~\cite{2015_Tranquada_high-Tc_intertwined_orders,
2019_Schmalian_vestigal_order_in_QM}.
Interestingly,
it was noted by Kivelson and Emery in 1996 that the phase diagrams of polyacetylene and the cuprates bear remarkable similarities~\cite{1996_Kivelson_topological_doping_polyacetylene} and the intertwined AFM and CDW/SDW orders have been stabilized before in the cuprates~\cite{2020_Yubo_SCAN_stripe_YBCO,
2024_Nokelainen_ybco_pressure}.
Our findings thus suggest that there could be further similarities between the physics of polyacetylene and the cuprates.


Finally,
we note that DFT provides a simplified picture of the real system:
every XC functional is only an approximation and has its own flaws.
For example,
PBE is likely to underestimate the correlation,
while SCAN can overestimate it in some cases such as the elemental transition metals~\cite{2019_Trickey_SCAN_TM_overmagnetization,
2020_Singh_SCAN_magnetism_shortcomings}.
Furthermore,
DFT is strictly a static zero-temperature theory for the ground state,
so it does not include for example important dynamical effects such as the quantum zero-point motion that could affect the stability of the CDW state~\cite{2013_Hudson_bond_alternation,2018_Hudson_polyacetylene_review} and introduce corrections beyond the band theory,
including the electron-phonon coupling effects~\cite{2011_Cannuccia_polyacetylene_zero_point_motion}.
Therefore,
drawing definite conclusions based on the DFT results can be difficult,
but like in the case of the cuprates,
DFT and all the different XC descriptions can at the very least provide valuable \quotemarks{snapshots} of the time-dependent solution space and the complicated phenomena at play in the real material.

\section{Conclusions}

We discuss the role of electronic correlation effects in polyacetylene using the DFT,
based on various exchange-correlation functionals across the Jacob's ladder framework.
We find that correlation can stabilize a SDW state with antiferromagnetically ordered C magnetic moments with the GGA, meta-GGA and hybrid functionals,
which to our knowledge has not been reported previously in the \emph{ab initio} level.
The B3LYP hybrid functional predicts that the charge density wave (CDW) state with bond alternation in accord with earlier studies.
The conventional understanding has been that the CDW is the ground state,
although this has been disputed by Hudson~\cite{2013_Hudson_bond_alternation,2018_Hudson_polyacetylene_review}.
The persistence of the SDW state found with various functionals indicates that the actual ground state may be far more complex,
resembling the fluctuating stripe phases observed in cuprates~\cite{2020_Yubo_SCAN_stripe_YBCO}.
Testing this scenario in polyacetylene would require simulations with larger computational unit cells that can accommodate stripe phases with larger periodicity.
A variety of orders driven by the charge, spin and lattice degrees of freedom could also compete to produce a pseudogap like state of the cuprates and contribute to the observed band gap in polyacetylene.
Our study thus further highlights the similarities between polyacetylene and the cuprates,
which have been pointed to by Kivelson~\cite{1996_Kivelson_topological_doping_polyacetylene}.
Our findings not only contribute to a deeper understanding of this foundational organic polymer but also emphasize the need for comprehensive \emph{ab initio} studies to uncover the interplay of electronic and structural factors in one-dimensional carbon-based materials,
potentially opening new avenues for the application of polyacetylene in organic magnetism and microelectronics.

\section*{Acknowledgments}

The work at LUT was supported by the INERCOM platform.
The authors acknowledge Northeastern University’s Advanced Scientific Computation Center and the Discovery Cluster for computational resources.
The work at Northeastern benefited from the Massachusetts Technology Collaborative through award number 22032 and the resources of the Quantum Materials and Sensing Institute.


\begin{thebibliography}{55}%
\makeatletter
\providecommand \@ifxundefined [1]{%
 \@ifx{#1\undefined}
}%
\providecommand \@ifnum [1]{%
 \ifnum #1\expandafter \@firstoftwo
 \else \expandafter \@secondoftwo
 \fi
}%
\providecommand \@ifx [1]{%
 \ifx #1\expandafter \@firstoftwo
 \else \expandafter \@secondoftwo
 \fi
}%
\providecommand \natexlab [1]{#1}%
\providecommand \enquote  [1]{``#1''}%
\providecommand \bibnamefont  [1]{#1}%
\providecommand \bibfnamefont [1]{#1}%
\providecommand \citenamefont [1]{#1}%
\providecommand \href@noop [0]{\@secondoftwo}%
\providecommand \href [0]{\begingroup \@sanitize@url \@href}%
\providecommand \@href[1]{\@@startlink{#1}\@@href}%
\providecommand \@@href[1]{\endgroup#1\@@endlink}%
\providecommand \@sanitize@url [0]{\catcode `\\12\catcode `\$12\catcode `\&12\catcode `\#12\catcode `\^12\catcode `\_12\catcode `\%12\relax}%
\providecommand \@@startlink[1]{}%
\providecommand \@@endlink[0]{}%
\providecommand \url  [0]{\begingroup\@sanitize@url \@url }%
\providecommand \@url [1]{\endgroup\@href {#1}{\urlprefix }}%
\providecommand \urlprefix  [0]{URL }%
\providecommand \Eprint [0]{\href }%
\providecommand \doibase [0]{http://dx.doi.org/}%
\providecommand \selectlanguage [0]{\@gobble}%
\providecommand \bibinfo  [0]{\@secondoftwo}%
\providecommand \bibfield  [0]{\@secondoftwo}%
\providecommand \translation [1]{[#1]}%
\providecommand \BibitemOpen [0]{}%
\providecommand \bibitemStop [0]{}%
\providecommand \bibitemNoStop [0]{.\EOS\space}%
\providecommand \EOS [0]{\spacefactor3000\relax}%
\providecommand \BibitemShut  [1]{\csname bibitem#1\endcsname}%
\let\auto@bib@innerbib\@empty
\bibitem [{\citenamefont {Baeriswyl}\ and\ \citenamefont {Campbell}(1992)}]{1992_Kiess_polymer_book}%
  \BibitemOpen
  \bibfield  {author} {\bibinfo {author} {\bibfnamefont {D.}~\bibnamefont {Baeriswyl}}\ and\ \bibinfo {author} {\bibfnamefont {D.}~\bibnamefont {Campbell}},\ }\href {\doibase https://doi.org/10.1007/978-3-642-46729-5} {\emph {\bibinfo {title} {Conjugated Conducting Polymers}}},\ edited by\ \bibinfo {editor} {\bibfnamefont {H.~G.}\ \bibnamefont {Kiess}}\ (\bibinfo  {publisher} {Springer-Verlag},\ \bibinfo {year} {1992})\BibitemShut {NoStop}%
\bibitem [{\citenamefont {Barford}(2005)}]{2005_Barford_polymer_book}%
  \BibitemOpen
  \bibfield  {author} {\bibinfo {author} {\bibfnamefont {W.}~\bibnamefont {Barford}},\ }\href@noop {} {\emph {\bibinfo {title} {Electronic and Optical Properties of Conjugated Polymers}}},\ Vol.\ \bibinfo {volume} {129}\ (\bibinfo  {publisher} {Oxford University Press},\ \bibinfo {year} {2005})\BibitemShut {NoStop}%
\bibitem [{\citenamefont {Barbiellini}\ \emph {et~al.}(1985)\citenamefont {Barbiellini}, \citenamefont {Buksh}, \citenamefont {Cecchet}, \citenamefont {Hemery}, \citenamefont {Lemeilleur}, \citenamefont {Rancoita}, \citenamefont {Vismara},\ and\ \citenamefont {Seidman}}]{Barbiellini_1985}%
  \BibitemOpen
  \bibfield  {author} {\bibinfo {author} {\bibfnamefont {G.}~\bibnamefont {Barbiellini}}, \bibinfo {author} {\bibfnamefont {P.}~\bibnamefont {Buksh}}, \bibinfo {author} {\bibfnamefont {G.}~\bibnamefont {Cecchet}}, \bibinfo {author} {\bibfnamefont {J.}~\bibnamefont {Hemery}}, \bibinfo {author} {\bibfnamefont {F.}~\bibnamefont {Lemeilleur}}, \bibinfo {author} {\bibfnamefont {P.}~\bibnamefont {Rancoita}}, \bibinfo {author} {\bibfnamefont {G.}~\bibnamefont {Vismara}}, \ and\ \bibinfo {author} {\bibfnamefont {A.}~\bibnamefont {Seidman}},\ }\href {\doibase https://doi.org/10.1016/0168-9002(85)90556-X} {\bibfield  {journal} {\bibinfo  {journal} {Nuclear Instruments and Methods in Physics Research Section A: Accelerators, Spectrometers, Detectors and Associated Equipment}\ }\textbf {\bibinfo {volume} {235}},\ \bibinfo {pages} {216} (\bibinfo {year} {1985})}\BibitemShut {NoStop}%
\bibitem [{\citenamefont {Beckerle}\ and\ \citenamefont {Ströbele}(2000)}]{beckerle_2000}%
  \BibitemOpen
  \bibfield  {author} {\bibinfo {author} {\bibfnamefont {P.}~\bibnamefont {Beckerle}}\ and\ \bibinfo {author} {\bibfnamefont {H.}~\bibnamefont {Ströbele}},\ }\href {\doibase https://doi.org/10.1016/S0168-9002(00)00255-2} {\bibfield  {journal} {\bibinfo  {journal} {Nuclear Instruments and Methods in Physics Research Section A: Accelerators, Spectrometers, Detectors and Associated Equipment}\ }\textbf {\bibinfo {volume} {449}},\ \bibinfo {pages} {302} (\bibinfo {year} {2000})}\BibitemShut {NoStop}%
\bibitem [{\citenamefont {Chiang}\ \emph {et~al.}(1977)\citenamefont {Chiang}, \citenamefont {Fincher}, \citenamefont {Park}, \citenamefont {Heeger}, \citenamefont {Shirakawa}, \citenamefont {Louis}, \citenamefont {Gau},\ and\ \citenamefont {MacDiarmid}}]{1977_Chiang_polyacetylene_conductivity}%
  \BibitemOpen
  \bibfield  {author} {\bibinfo {author} {\bibfnamefont {C.~K.}\ \bibnamefont {Chiang}}, \bibinfo {author} {\bibfnamefont {C.~R.}\ \bibnamefont {Fincher}}, \bibinfo {author} {\bibfnamefont {Y.~W.}\ \bibnamefont {Park}}, \bibinfo {author} {\bibfnamefont {A.~J.}\ \bibnamefont {Heeger}}, \bibinfo {author} {\bibfnamefont {H.}~\bibnamefont {Shirakawa}}, \bibinfo {author} {\bibfnamefont {E.~J.}\ \bibnamefont {Louis}}, \bibinfo {author} {\bibfnamefont {S.~C.}\ \bibnamefont {Gau}}, \ and\ \bibinfo {author} {\bibfnamefont {A.~G.}\ \bibnamefont {MacDiarmid}},\ }\href {\doibase 10.1103/PhysRevLett.39.1098} {\bibfield  {journal} {\bibinfo  {journal} {Phys. Rev. Lett.}\ }\textbf {\bibinfo {volume} {39}},\ \bibinfo {pages} {1098} (\bibinfo {year} {1977})}\BibitemShut {NoStop}%
\bibitem [{\citenamefont {Shirakawa}(2001)}]{2001_Shirakawa_polyacetylene_nobel_lecture}%
  \BibitemOpen
  \bibfield  {author} {\bibinfo {author} {\bibfnamefont {H.}~\bibnamefont {Shirakawa}},\ }\href {\doibase 10.1103/RevModPhys.73.713} {\bibfield  {journal} {\bibinfo  {journal} {Rev. Mod. Phys.}\ }\textbf {\bibinfo {volume} {73}},\ \bibinfo {pages} {713} (\bibinfo {year} {2001})}\BibitemShut {NoStop}%
\bibitem [{\citenamefont {Wang}\ \emph {et~al.}(2019)\citenamefont {Wang}, \citenamefont {Sun}, \citenamefont {Gr{\"o}ning}, \citenamefont {Widmer}, \citenamefont {Pignedoli}, \citenamefont {Cai}, \citenamefont {Yu}, \citenamefont {Yuan}, \citenamefont {Li}, \citenamefont {Ju}, \citenamefont {Zhu}, \citenamefont {Ruffieux}, \citenamefont {Fasel},\ and\ \citenamefont {Xu}}]{2019_Wang_polyacetylene_EXP}%
  \BibitemOpen
  \bibfield  {author} {\bibinfo {author} {\bibfnamefont {S.}~\bibnamefont {Wang}}, \bibinfo {author} {\bibfnamefont {Q.}~\bibnamefont {Sun}}, \bibinfo {author} {\bibfnamefont {O.}~\bibnamefont {Gr{\"o}ning}}, \bibinfo {author} {\bibfnamefont {R.}~\bibnamefont {Widmer}}, \bibinfo {author} {\bibfnamefont {C.~A.}\ \bibnamefont {Pignedoli}}, \bibinfo {author} {\bibfnamefont {L.}~\bibnamefont {Cai}}, \bibinfo {author} {\bibfnamefont {X.}~\bibnamefont {Yu}}, \bibinfo {author} {\bibfnamefont {B.}~\bibnamefont {Yuan}}, \bibinfo {author} {\bibfnamefont {C.}~\bibnamefont {Li}}, \bibinfo {author} {\bibfnamefont {H.}~\bibnamefont {Ju}}, \bibinfo {author} {\bibfnamefont {J.}~\bibnamefont {Zhu}}, \bibinfo {author} {\bibfnamefont {P.}~\bibnamefont {Ruffieux}}, \bibinfo {author} {\bibfnamefont {R.}~\bibnamefont {Fasel}}, \ and\ \bibinfo {author} {\bibfnamefont {W.}~\bibnamefont {Xu}},\ }\href@noop {} {\bibfield  {journal} {\bibinfo  {journal} {Nat. Chem.}\ }\textbf {\bibinfo {volume} {11}},\ \bibinfo {pages} {924} (\bibinfo
  {year} {2019})}\BibitemShut {NoStop}%
\bibitem [{\citenamefont {Su}\ \emph {et~al.}(1979)\citenamefont {Su}, \citenamefont {Schrieffer},\ and\ \citenamefont {Heeger}}]{1979_SSH}%
  \BibitemOpen
  \bibfield  {author} {\bibinfo {author} {\bibfnamefont {W.~P.}\ \bibnamefont {Su}}, \bibinfo {author} {\bibfnamefont {J.~R.}\ \bibnamefont {Schrieffer}}, \ and\ \bibinfo {author} {\bibfnamefont {A.~J.}\ \bibnamefont {Heeger}},\ }\href {\doibase 10.1103/PhysRevLett.42.1698} {\bibfield  {journal} {\bibinfo  {journal} {Phys. Rev. Lett.}\ }\textbf {\bibinfo {volume} {42}},\ \bibinfo {pages} {1698} (\bibinfo {year} {1979})}\BibitemShut {NoStop}%
\bibitem [{\citenamefont {Rice}(1979)}]{1979_Rice_solitons_polyacetylene}%
  \BibitemOpen
  \bibfield  {author} {\bibinfo {author} {\bibfnamefont {M.~J.}\ \bibnamefont {Rice}},\ }\href {\doibase https://doi.org/10.1016/0375-9601(79)90905-8} {\bibfield  {journal} {\bibinfo  {journal} {Phys. Lett. A}\ }\textbf {\bibinfo {volume} {71}},\ \bibinfo {pages} {152} (\bibinfo {year} {1979})}\BibitemShut {NoStop}%
\bibitem [{\citenamefont {Ramasesha}\ \emph {et~al.}(2000)\citenamefont {Ramasesha}, \citenamefont {Pati}, \citenamefont {Shuai},\ and\ \citenamefont {Br{\'e}das}}]{2000_Ramasesha_review_conjugated_polymers_density_matrix_renormalization}%
  \BibitemOpen
  \bibfield  {author} {\bibinfo {author} {\bibfnamefont {S.}~\bibnamefont {Ramasesha}}, \bibinfo {author} {\bibfnamefont {S.~K.}\ \bibnamefont {Pati}}, \bibinfo {author} {\bibfnamefont {Z.}~\bibnamefont {Shuai}}, \ and\ \bibinfo {author} {\bibfnamefont {J.}~\bibnamefont {Br{\'e}das}},\ }\href {\doibase https://doi.org/10.1016/S0065-3276(00)38004-2} {\ \bibinfo {series} {Adv. Quantum Chem.},\ \textbf {\bibinfo {volume} {38}},\ \bibinfo {pages} {121} (\bibinfo {year} {2000})}\BibitemShut {NoStop}%
\bibitem [{\citenamefont {Champagne}\ \emph {et~al.}(1998)\citenamefont {Champagne}, \citenamefont {Perp{\'e}te}, \citenamefont {van Gisbergen}, \citenamefont {Baerends}, \citenamefont {Snijders}, \citenamefont {Soubra-Ghaoui}, \citenamefont {Robins},\ and\ \citenamefont {Kirtman}}]{1998_Champagne_polyacetylene_HF_DFT}%
  \BibitemOpen
  \bibfield  {author} {\bibinfo {author} {\bibfnamefont {B.}~\bibnamefont {Champagne}}, \bibinfo {author} {\bibfnamefont {E.~A.}\ \bibnamefont {Perp{\'e}te}}, \bibinfo {author} {\bibfnamefont {S.~J.~A.}\ \bibnamefont {van Gisbergen}}, \bibinfo {author} {\bibfnamefont {E.-J.}\ \bibnamefont {Baerends}}, \bibinfo {author} {\bibfnamefont {J.~G.}\ \bibnamefont {Snijders}}, \bibinfo {author} {\bibfnamefont {C.}~\bibnamefont {Soubra-Ghaoui}}, \bibinfo {author} {\bibfnamefont {K.~A.}\ \bibnamefont {Robins}}, \ and\ \bibinfo {author} {\bibfnamefont {B.}~\bibnamefont {Kirtman}},\ }\href {\doibase 10.1063/1.477731} {\bibfield  {journal} {\bibinfo  {journal} {J. Chem. Phys.}\ }\textbf {\bibinfo {volume} {109}},\ \bibinfo {pages} {10489} (\bibinfo {year} {1998})}\BibitemShut {NoStop}%
\bibitem [{\citenamefont {Ho~Choi}\ \emph {et~al.}(1997)\citenamefont {Ho~Choi}, \citenamefont {Kertesz},\ and\ \citenamefont {Karpfen}}]{1998_Choi_polyacetylene_correlation_BLA}%
  \BibitemOpen
  \bibfield  {author} {\bibinfo {author} {\bibfnamefont {C.}~\bibnamefont {Ho~Choi}}, \bibinfo {author} {\bibfnamefont {M.}~\bibnamefont {Kertesz}}, \ and\ \bibinfo {author} {\bibfnamefont {A.}~\bibnamefont {Karpfen}},\ }\href {\doibase 10.1063/1.474914} {\bibfield  {journal} {\bibinfo  {journal} {J. Chem. Phys.}\ }\textbf {\bibinfo {volume} {107}},\ \bibinfo {pages} {6712} (\bibinfo {year} {1997})}\BibitemShut {NoStop}%
\bibitem [{\citenamefont {Lacivita}\ \emph {et~al.}(2012)\citenamefont {Lacivita}, \citenamefont {R{\`e}rat}, \citenamefont {Orlando}, \citenamefont {Ferrero},\ and\ \citenamefont {Dovesi}}]{2012_Lacivita_polyacetylene_DFT_polarizability}%
  \BibitemOpen
  \bibfield  {author} {\bibinfo {author} {\bibfnamefont {V.}~\bibnamefont {Lacivita}}, \bibinfo {author} {\bibfnamefont {M.}~\bibnamefont {R{\`e}rat}}, \bibinfo {author} {\bibfnamefont {R.}~\bibnamefont {Orlando}}, \bibinfo {author} {\bibfnamefont {M.}~\bibnamefont {Ferrero}}, \ and\ \bibinfo {author} {\bibfnamefont {R.}~\bibnamefont {Dovesi}},\ }\href {\doibase 10.1063/1.3690457} {\bibfield  {journal} {\bibinfo  {journal} {J. Chem. Phys.}\ }\textbf {\bibinfo {volume} {136}},\ \bibinfo {pages} {114101} (\bibinfo {year} {2012})}\BibitemShut {NoStop}%
\bibitem [{\citenamefont {Chattopadhyaya}\ \emph {et~al.}(2014)\citenamefont {Chattopadhyaya}, \citenamefont {Sen}, \citenamefont {Alam},\ and\ \citenamefont {Chakrabarti}}]{2014_Chattopadhyaya_polymer_DFT_band_gap_correlation}%
  \BibitemOpen
  \bibfield  {author} {\bibinfo {author} {\bibfnamefont {M.}~\bibnamefont {Chattopadhyaya}}, \bibinfo {author} {\bibfnamefont {S.}~\bibnamefont {Sen}}, \bibinfo {author} {\bibfnamefont {M.}~\bibnamefont {Alam}}, \ and\ \bibinfo {author} {\bibfnamefont {S.}~\bibnamefont {Chakrabarti}},\ }\href {\doibase https://doi.org/10.1016/j.jpcs.2013.09.018} {\bibfield  {journal} {\bibinfo  {journal} {J. Phys. Chem. Solids}\ }\textbf {\bibinfo {volume} {75}},\ \bibinfo {pages} {212} (\bibinfo {year} {2014})}\BibitemShut {NoStop}%
\bibitem [{\citenamefont {Barborini}\ and\ \citenamefont {Guidoni}(2015)}]{2015_Barborini_polyacetylene_QMC_RVB_hybrid_functionals}%
  \BibitemOpen
  \bibfield  {author} {\bibinfo {author} {\bibfnamefont {M.}~\bibnamefont {Barborini}}\ and\ \bibinfo {author} {\bibfnamefont {L.}~\bibnamefont {Guidoni}},\ }\href {\doibase 10.1021/acs.jctc.5b00427} {\bibfield  {journal} {\bibinfo  {journal} {J. Chem. Theory Comput.}\ }\textbf {\bibinfo {volume} {11}},\ \bibinfo {pages} {4109} (\bibinfo {year} {2015})}\BibitemShut {NoStop}%
\bibitem [{\citenamefont {Heeger}\ \emph {et~al.}(1988)\citenamefont {Heeger}, \citenamefont {Kivelson}, \citenamefont {Schrieffer},\ and\ \citenamefont {Su}}]{1988_Heeger_review_soliton_polyacetylene}%
  \BibitemOpen
  \bibfield  {author} {\bibinfo {author} {\bibfnamefont {A.~J.}\ \bibnamefont {Heeger}}, \bibinfo {author} {\bibfnamefont {S.}~\bibnamefont {Kivelson}}, \bibinfo {author} {\bibfnamefont {J.~R.}\ \bibnamefont {Schrieffer}}, \ and\ \bibinfo {author} {\bibfnamefont {W.~P.}\ \bibnamefont {Su}},\ }\href {\doibase 10.1103/RevModPhys.60.781} {\bibfield  {journal} {\bibinfo  {journal} {Rev. Mod. Phys.}\ }\textbf {\bibinfo {volume} {60}},\ \bibinfo {pages} {781} (\bibinfo {year} {1988})}\BibitemShut {NoStop}%
\bibitem [{\citenamefont {Soos}\ \emph {et~al.}(1994)\citenamefont {Soos}, \citenamefont {Galvão},\ and\ \citenamefont {Etemad}}]{1994_Soos_polymer_fluorescene_structure}%
  \BibitemOpen
  \bibfield  {author} {\bibinfo {author} {\bibfnamefont {Z.~G.}\ \bibnamefont {Soos}}, \bibinfo {author} {\bibfnamefont {D.~S.}\ \bibnamefont {Galvão}}, \ and\ \bibinfo {author} {\bibfnamefont {S.}~\bibnamefont {Etemad}},\ }\href {\doibase https://doi.org/10.1002/adma.19940060404} {\bibfield  {journal} {\bibinfo  {journal} {Adv. Mater.}\ }\textbf {\bibinfo {volume} {6}},\ \bibinfo {pages} {280} (\bibinfo {year} {1994})}\BibitemShut {NoStop}%
\bibitem [{\citenamefont {Hudson}\ and\ \citenamefont {Allis}(2013)}]{2013_Hudson_bond_alternation}%
  \BibitemOpen
  \bibfield  {author} {\bibinfo {author} {\bibfnamefont {B.~S.}\ \bibnamefont {Hudson}}\ and\ \bibinfo {author} {\bibfnamefont {D.~G.}\ \bibnamefont {Allis}},\ }\href {\doibase 10.1016/j.molstruc.2012.07.051} {\bibfield  {journal} {\bibinfo  {journal} {J. Mol. Struct.}\ }\textbf {\bibinfo {volume} {1032}},\ \bibinfo {pages} {78} (\bibinfo {year} {2013})}\BibitemShut {NoStop}%
\bibitem [{\citenamefont {Hudson}(2018)}]{2018_Hudson_polyacetylene_review}%
  \BibitemOpen
  \bibfield  {author} {\bibinfo {author} {\bibfnamefont {B.~S.}\ \bibnamefont {Hudson}},\ }\href {\doibase 10.3390/ma11020242} {\bibfield  {journal} {\bibinfo  {journal} {Materials}\ }\textbf {\bibinfo {volume} {11}} (\bibinfo {year} {2018}),\ 10.3390/ma11020242}\BibitemShut {NoStop}%
\bibitem [{\citenamefont {Harris}\ and\ \citenamefont {Falicov}(1969)}]{1969_Harris_polyacetylene_charge_spin_density_waves}%
  \BibitemOpen
  \bibfield  {author} {\bibinfo {author} {\bibfnamefont {R.~A.}\ \bibnamefont {Harris}}\ and\ \bibinfo {author} {\bibfnamefont {L.~M.}\ \bibnamefont {Falicov}},\ }\href {\doibase 10.1063/1.1671900} {\bibfield  {journal} {\bibinfo  {journal} {J. Chem. Phys.}\ }\textbf {\bibinfo {volume} {51}},\ \bibinfo {pages} {5034} (\bibinfo {year} {1969})}\BibitemShut {NoStop}%
\bibitem [{\citenamefont {Subbaswamy}\ and\ \citenamefont {Grabowski}(1981)}]{1981_Subbaswamy_polyacetylene_correlation}%
  \BibitemOpen
  \bibfield  {author} {\bibinfo {author} {\bibfnamefont {K.~R.}\ \bibnamefont {Subbaswamy}}\ and\ \bibinfo {author} {\bibfnamefont {M.}~\bibnamefont {Grabowski}},\ }\href {\doibase 10.1103/PhysRevB.24.2168} {\bibfield  {journal} {\bibinfo  {journal} {Phys. Rev. B}\ }\textbf {\bibinfo {volume} {24}},\ \bibinfo {pages} {2168} (\bibinfo {year} {1981})}\BibitemShut {NoStop}%
\bibitem [{\citenamefont {Fukutome}\ and\ \citenamefont {Sasai}(1982)}]{1982_Fukutome_polyacetylene_charge_spin_density_waves}%
  \BibitemOpen
  \bibfield  {author} {\bibinfo {author} {\bibfnamefont {H.}~\bibnamefont {Fukutome}}\ and\ \bibinfo {author} {\bibfnamefont {M.}~\bibnamefont {Sasai}},\ }\href {\doibase 10.1143/PTP.67.41} {\bibfield  {journal} {\bibinfo  {journal} {Prog. Theor. Phys.}\ }\textbf {\bibinfo {volume} {67}},\ \bibinfo {pages} {41} (\bibinfo {year} {1982})}\BibitemShut {NoStop}%
\bibitem [{\citenamefont {Lee}\ \emph {et~al.}(2006)\citenamefont {Lee}, \citenamefont {Nagaosa},\ and\ \citenamefont {Wen}}]{2006_Lee_review_high-Tc_physics}%
  \BibitemOpen
  \bibfield  {author} {\bibinfo {author} {\bibfnamefont {P.~A.}\ \bibnamefont {Lee}}, \bibinfo {author} {\bibfnamefont {N.}~\bibnamefont {Nagaosa}}, \ and\ \bibinfo {author} {\bibfnamefont {X.-G.}\ \bibnamefont {Wen}},\ }\href {\doibase 10.1103/RevModPhys.78.17} {\bibfield  {journal} {\bibinfo  {journal} {Rev. Mod. Phys.}\ }\textbf {\bibinfo {volume} {78}},\ \bibinfo {pages} {17} (\bibinfo {year} {2006})}\BibitemShut {NoStop}%
\bibitem [{\citenamefont {Horsch}(1981)}]{1981_Horsch_polyacetylene_correlation_BLA}%
  \BibitemOpen
  \bibfield  {author} {\bibinfo {author} {\bibfnamefont {P.}~\bibnamefont {Horsch}},\ }\href {\doibase 10.1103/PhysRevB.24.7351} {\bibfield  {journal} {\bibinfo  {journal} {Phys. Rev. B}\ }\textbf {\bibinfo {volume} {24}},\ \bibinfo {pages} {7351} (\bibinfo {year} {1981})}\BibitemShut {NoStop}%
\bibitem [{\citenamefont {Baeriswyl}\ and\ \citenamefont {Maki}(1985)}]{1985_Baeriswyl_polyacetylene_correlation}%
  \BibitemOpen
  \bibfield  {author} {\bibinfo {author} {\bibfnamefont {D.}~\bibnamefont {Baeriswyl}}\ and\ \bibinfo {author} {\bibfnamefont {K.}~\bibnamefont {Maki}},\ }\href {\doibase 10.1103/PhysRevB.31.6633} {\bibfield  {journal} {\bibinfo  {journal} {Phys. Rev. B}\ }\textbf {\bibinfo {volume} {31}},\ \bibinfo {pages} {6633} (\bibinfo {year} {1985})}\BibitemShut {NoStop}%
\bibitem [{\citenamefont {Ma}\ \emph {et~al.}(2002)\citenamefont {Ma}, \citenamefont {Li},\ and\ \citenamefont {Jiang}}]{2002_Ma_polyacetylene_time_dependent_DFT}%
  \BibitemOpen
  \bibfield  {author} {\bibinfo {author} {\bibfnamefont {J.}~\bibnamefont {Ma}}, \bibinfo {author} {\bibfnamefont {S.}~\bibnamefont {Li}}, \ and\ \bibinfo {author} {\bibfnamefont {Y.}~\bibnamefont {Jiang}},\ }\href {\doibase 10.1021/ma011279m} {\bibfield  {journal} {\bibinfo  {journal} {Macromolecules}\ }\textbf {\bibinfo {volume} {35}},\ \bibinfo {pages} {1109} (\bibinfo {year} {2002})}\BibitemShut {NoStop}%
\bibitem [{\citenamefont {Furuya}\ \emph {et~al.}(2023)\citenamefont {Furuya}, \citenamefont {Sakamoto},\ and\ \citenamefont {Tasumi}}]{2023_Furuya_polyacetylene_DFT_vibrations}%
  \BibitemOpen
  \bibfield  {author} {\bibinfo {author} {\bibfnamefont {K.}~\bibnamefont {Furuya}}, \bibinfo {author} {\bibfnamefont {A.}~\bibnamefont {Sakamoto}}, \ and\ \bibinfo {author} {\bibfnamefont {M.}~\bibnamefont {Tasumi}},\ }\href {\doibase 10.1021/acs.jpca.3c02180} {\bibfield  {journal} {\bibinfo  {journal} {J. Phys. Chem. A}\ }\textbf {\bibinfo {volume} {127}},\ \bibinfo {pages} {5344} (\bibinfo {year} {2023})}\BibitemShut {NoStop}%
\bibitem [{\citenamefont {Winter}\ \emph {et~al.}(1990)\citenamefont {Winter}, \citenamefont {Sachs}, \citenamefont {Dormann}, \citenamefont {Cosmo},\ and\ \citenamefont {Naarmann}}]{1990_Winter_polyacetylene_free_radicals_magnetism}%
  \BibitemOpen
  \bibfield  {author} {\bibinfo {author} {\bibfnamefont {H.}~\bibnamefont {Winter}}, \bibinfo {author} {\bibfnamefont {G.}~\bibnamefont {Sachs}}, \bibinfo {author} {\bibfnamefont {E.}~\bibnamefont {Dormann}}, \bibinfo {author} {\bibfnamefont {R.}~\bibnamefont {Cosmo}}, \ and\ \bibinfo {author} {\bibfnamefont {H.}~\bibnamefont {Naarmann}},\ }\href {\doibase https://doi.org/10.1016/0379-6779(90)90259-N} {\bibfield  {journal} {\bibinfo  {journal} {Synth. Met.}\ }\textbf {\bibinfo {volume} {36}},\ \bibinfo {pages} {353} (\bibinfo {year} {1990})}\BibitemShut {NoStop}%
\bibitem [{\citenamefont {Cosmo}\ \emph {et~al.}(1991)\citenamefont {Cosmo}, \citenamefont {Dormann}, \citenamefont {Gotschy}, \citenamefont {Naarmann},\ and\ \citenamefont {Winter}}]{1991_Cosmo_polyacetylene_free_radicals_magnetism}%
  \BibitemOpen
  \bibfield  {author} {\bibinfo {author} {\bibfnamefont {R.}~\bibnamefont {Cosmo}}, \bibinfo {author} {\bibfnamefont {E.}~\bibnamefont {Dormann}}, \bibinfo {author} {\bibfnamefont {B.}~\bibnamefont {Gotschy}}, \bibinfo {author} {\bibfnamefont {H.}~\bibnamefont {Naarmann}}, \ and\ \bibinfo {author} {\bibfnamefont {H.}~\bibnamefont {Winter}},\ }\href {\doibase https://doi.org/10.1016/0379-6779(91)91084-N} {\bibfield  {journal} {\bibinfo  {journal} {Synth. Met.}\ }\textbf {\bibinfo {volume} {41}},\ \bibinfo {pages} {369} (\bibinfo {year} {1991})},\ \bibinfo {note} {proceedings of the International Conference on Science and Technology of Synthetic Metals}\BibitemShut {NoStop}%
\bibitem [{\citenamefont {Perdew}\ and\ \citenamefont {Schmidt}(2001)}]{2001_Perdew_DFT_Jacobs_ladder}%
  \BibitemOpen
  \bibfield  {author} {\bibinfo {author} {\bibfnamefont {J.~P.}\ \bibnamefont {Perdew}}\ and\ \bibinfo {author} {\bibfnamefont {K.}~\bibnamefont {Schmidt}},\ }\href {\doibase 10.1063/1.1390175} {\bibfield  {journal} {\bibinfo  {journal} {AIP Conference Proceedings}\ }\textbf {\bibinfo {volume} {577}},\ \bibinfo {pages} {1} (\bibinfo {year} {2001})}\BibitemShut {NoStop}%
\bibitem [{\citenamefont {Perdew}\ \emph {et~al.}(1996)\citenamefont {Perdew}, \citenamefont {Burke},\ and\ \citenamefont {Ernzerhof}}]{1996_PBE_functional}%
  \BibitemOpen
  \bibfield  {author} {\bibinfo {author} {\bibfnamefont {J.~P.}\ \bibnamefont {Perdew}}, \bibinfo {author} {\bibfnamefont {K.}~\bibnamefont {Burke}}, \ and\ \bibinfo {author} {\bibfnamefont {M.}~\bibnamefont {Ernzerhof}},\ }\href {\doibase 10.1103/PhysRevLett.77.3865} {\bibfield  {journal} {\bibinfo  {journal} {Phys. Rev. Lett.}\ }\textbf {\bibinfo {volume} {77}},\ \bibinfo {pages} {3865} (\bibinfo {year} {1996})}\BibitemShut {NoStop}%
\bibitem [{\citenamefont {Sun}\ \emph {et~al.}(2015)\citenamefont {Sun}, \citenamefont {Ruzsinszky},\ and\ \citenamefont {Perdew}}]{2015_Sun_SCAN}%
  \BibitemOpen
  \bibfield  {author} {\bibinfo {author} {\bibfnamefont {J.}~\bibnamefont {Sun}}, \bibinfo {author} {\bibfnamefont {A.}~\bibnamefont {Ruzsinszky}}, \ and\ \bibinfo {author} {\bibfnamefont {J.~P.}\ \bibnamefont {Perdew}},\ }\href {\doibase 10.1103/PhysRevLett.115.036402} {\bibfield  {journal} {\bibinfo  {journal} {Phys. Rev. Lett.}\ }\textbf {\bibinfo {volume} {115}},\ \bibinfo {pages} {036402} (\bibinfo {year} {2015})}\BibitemShut {NoStop}%
\bibitem [{\citenamefont {Lane}\ \emph {et~al.}(2018)\citenamefont {Lane}, \citenamefont {Furness}, \citenamefont {Buda}, \citenamefont {Zhang}, \citenamefont {Markiewicz}, \citenamefont {Barbiellini}, \citenamefont {Sun},\ and\ \citenamefont {Bansil}}]{2018_Chris_La2CuO4_SCAN}%
  \BibitemOpen
  \bibfield  {author} {\bibinfo {author} {\bibfnamefont {C.}~\bibnamefont {Lane}}, \bibinfo {author} {\bibfnamefont {J.~W.}\ \bibnamefont {Furness}}, \bibinfo {author} {\bibfnamefont {I.~G.}\ \bibnamefont {Buda}}, \bibinfo {author} {\bibfnamefont {Y.}~\bibnamefont {Zhang}}, \bibinfo {author} {\bibfnamefont {R.~S.}\ \bibnamefont {Markiewicz}}, \bibinfo {author} {\bibfnamefont {B.}~\bibnamefont {Barbiellini}}, \bibinfo {author} {\bibfnamefont {J.}~\bibnamefont {Sun}}, \ and\ \bibinfo {author} {\bibfnamefont {A.}~\bibnamefont {Bansil}},\ }\href {\doibase 10.1103/PhysRevB.98.125140} {\bibfield  {journal} {\bibinfo  {journal} {Phys. Rev. B}\ }\textbf {\bibinfo {volume} {98}},\ \bibinfo {pages} {125140} (\bibinfo {year} {2018})}\BibitemShut {NoStop}%
\bibitem [{\citenamefont {Isaacs}\ and\ \citenamefont {Wolverton}(2018)}]{2018_Isaacs_SCAN_performance_solids}%
  \BibitemOpen
  \bibfield  {author} {\bibinfo {author} {\bibfnamefont {E.~B.}\ \bibnamefont {Isaacs}}\ and\ \bibinfo {author} {\bibfnamefont {C.}~\bibnamefont {Wolverton}},\ }\href {\doibase 10.1103/PhysRevMaterials.2.063801} {\bibfield  {journal} {\bibinfo  {journal} {Phys. Rev. Materials}\ }\textbf {\bibinfo {volume} {2}},\ \bibinfo {pages} {063801} (\bibinfo {year} {2018})}\BibitemShut {NoStop}%
\bibitem [{\citenamefont {Nokelainen}\ \emph {et~al.}(2020)\citenamefont {Nokelainen}, \citenamefont {Lane}, \citenamefont {Markiewicz}, \citenamefont {Barbiellini}, \citenamefont {Pulkkinen}, \citenamefont {Singh}, \citenamefont {Sun}, \citenamefont {Pussi},\ and\ \citenamefont {Bansil}}]{2020_Nokelainen_bisco}%
  \BibitemOpen
  \bibfield  {author} {\bibinfo {author} {\bibfnamefont {J.}~\bibnamefont {Nokelainen}}, \bibinfo {author} {\bibfnamefont {C.}~\bibnamefont {Lane}}, \bibinfo {author} {\bibfnamefont {R.~S.}\ \bibnamefont {Markiewicz}}, \bibinfo {author} {\bibfnamefont {B.}~\bibnamefont {Barbiellini}}, \bibinfo {author} {\bibfnamefont {A.}~\bibnamefont {Pulkkinen}}, \bibinfo {author} {\bibfnamefont {B.}~\bibnamefont {Singh}}, \bibinfo {author} {\bibfnamefont {J.}~\bibnamefont {Sun}}, \bibinfo {author} {\bibfnamefont {K.}~\bibnamefont {Pussi}}, \ and\ \bibinfo {author} {\bibfnamefont {A.}~\bibnamefont {Bansil}},\ }\href {\doibase 10.1103/PhysRevB.101.214523} {\bibfield  {journal} {\bibinfo  {journal} {Phys. Rev. B}\ }\textbf {\bibinfo {volume} {101}},\ \bibinfo {pages} {214523} (\bibinfo {year} {2020})}\BibitemShut {NoStop}%
\bibitem [{\citenamefont {Zhang}\ \emph {et~al.}(2020)\citenamefont {Zhang}, \citenamefont {Lane}, \citenamefont {Furness}, \citenamefont {Barbiellini}, \citenamefont {Perdew}, \citenamefont {Markiewicz}, \citenamefont {Bansil},\ and\ \citenamefont {Sun}}]{2020_Yubo_SCAN_stripe_YBCO}%
  \BibitemOpen
  \bibfield  {author} {\bibinfo {author} {\bibfnamefont {Y.}~\bibnamefont {Zhang}}, \bibinfo {author} {\bibfnamefont {C.}~\bibnamefont {Lane}}, \bibinfo {author} {\bibfnamefont {J.~W.}\ \bibnamefont {Furness}}, \bibinfo {author} {\bibfnamefont {B.}~\bibnamefont {Barbiellini}}, \bibinfo {author} {\bibfnamefont {J.~P.}\ \bibnamefont {Perdew}}, \bibinfo {author} {\bibfnamefont {R.~S.}\ \bibnamefont {Markiewicz}}, \bibinfo {author} {\bibfnamefont {A.}~\bibnamefont {Bansil}}, \ and\ \bibinfo {author} {\bibfnamefont {J.}~\bibnamefont {Sun}},\ }\href {\doibase 10.1073/pnas.1910411116} {\bibfield  {journal} {\bibinfo  {journal} {Proc. Natl. Acad. Sci.}\ }\textbf {\bibinfo {volume} {117}},\ \bibinfo {pages} {68} (\bibinfo {year} {2020})}\BibitemShut {NoStop}%
\bibitem [{\citenamefont {Pokharel}\ \emph {et~al.}(2022)\citenamefont {Pokharel}, \citenamefont {Lane}, \citenamefont {Furness}, \citenamefont {Zhang}, \citenamefont {Ning}, \citenamefont {Barbiellini}, \citenamefont {Markiewicz}, \citenamefont {Zhang}, \citenamefont {Bansil},\ and\ \citenamefont {Sun}}]{2022_Kanun_functional_comparisons}%
  \BibitemOpen
  \bibfield  {author} {\bibinfo {author} {\bibfnamefont {K.}~\bibnamefont {Pokharel}}, \bibinfo {author} {\bibfnamefont {C.}~\bibnamefont {Lane}}, \bibinfo {author} {\bibfnamefont {J.~W.}\ \bibnamefont {Furness}}, \bibinfo {author} {\bibfnamefont {R.}~\bibnamefont {Zhang}}, \bibinfo {author} {\bibfnamefont {J.}~\bibnamefont {Ning}}, \bibinfo {author} {\bibfnamefont {B.}~\bibnamefont {Barbiellini}}, \bibinfo {author} {\bibfnamefont {R.~S.}\ \bibnamefont {Markiewicz}}, \bibinfo {author} {\bibfnamefont {Y.}~\bibnamefont {Zhang}}, \bibinfo {author} {\bibfnamefont {A.}~\bibnamefont {Bansil}}, \ and\ \bibinfo {author} {\bibfnamefont {J.}~\bibnamefont {Sun}},\ }\href {\doibase 10.1038/s41524-022-00711-z} {\bibfield  {journal} {\bibinfo  {journal} {Npj Comput. Mater.}\ }\textbf {\bibinfo {volume} {8}},\ \bibinfo {pages} {31} (\bibinfo {year} {2022})}\BibitemShut {NoStop}%
\bibitem [{\citenamefont {Nokelainen}\ \emph {et~al.}(2024)\citenamefont {Nokelainen}, \citenamefont {Matzelle}, \citenamefont {Lane}, \citenamefont {Atlam}, \citenamefont {Zhang}, \citenamefont {Markiewicz}, \citenamefont {Barbiellini}, \citenamefont {Sun},\ and\ \citenamefont {Bansil}}]{2024_Nokelainen_ybco_pressure}%
  \BibitemOpen
  \bibfield  {author} {\bibinfo {author} {\bibfnamefont {J.}~\bibnamefont {Nokelainen}}, \bibinfo {author} {\bibfnamefont {M.~E.}\ \bibnamefont {Matzelle}}, \bibinfo {author} {\bibfnamefont {C.}~\bibnamefont {Lane}}, \bibinfo {author} {\bibfnamefont {N.}~\bibnamefont {Atlam}}, \bibinfo {author} {\bibfnamefont {R.}~\bibnamefont {Zhang}}, \bibinfo {author} {\bibfnamefont {R.~S.}\ \bibnamefont {Markiewicz}}, \bibinfo {author} {\bibfnamefont {B.}~\bibnamefont {Barbiellini}}, \bibinfo {author} {\bibfnamefont {J.}~\bibnamefont {Sun}}, \ and\ \bibinfo {author} {\bibfnamefont {A.}~\bibnamefont {Bansil}},\ }\href {\doibase 10.1103/PhysRevB.110.L020502} {\bibfield  {journal} {\bibinfo  {journal} {Phys. Rev. B}\ }\textbf {\bibinfo {volume} {110}},\ \bibinfo {pages} {L020502} (\bibinfo {year} {2024})}\BibitemShut {NoStop}%
\bibitem [{\citenamefont {Zhang}\ \emph {et~al.}(2024)\citenamefont {Zhang}, \citenamefont {Lane}, \citenamefont {Nokelainen}, \citenamefont {Singh}, \citenamefont {Barbiellini}, \citenamefont {Markiewicz}, \citenamefont {Bansil},\ and\ \citenamefont {Sun}}]{2024_Ruiqi_nickelate_nemacity}%
  \BibitemOpen
  \bibfield  {author} {\bibinfo {author} {\bibfnamefont {R.}~\bibnamefont {Zhang}}, \bibinfo {author} {\bibfnamefont {C.}~\bibnamefont {Lane}}, \bibinfo {author} {\bibfnamefont {J.}~\bibnamefont {Nokelainen}}, \bibinfo {author} {\bibfnamefont {B.}~\bibnamefont {Singh}}, \bibinfo {author} {\bibfnamefont {B.}~\bibnamefont {Barbiellini}}, \bibinfo {author} {\bibfnamefont {R.~S.}\ \bibnamefont {Markiewicz}}, \bibinfo {author} {\bibfnamefont {A.}~\bibnamefont {Bansil}}, \ and\ \bibinfo {author} {\bibfnamefont {J.}~\bibnamefont {Sun}},\ }\href {\doibase 10.1103/PhysRevLett.133.066401} {\bibfield  {journal} {\bibinfo  {journal} {Phys. Rev. Lett.}\ }\textbf {\bibinfo {volume} {133}},\ \bibinfo {pages} {066401} (\bibinfo {year} {2024})}\BibitemShut {NoStop}%
\bibitem [{\citenamefont {Furness}\ \emph {et~al.}(2020)\citenamefont {Furness}, \citenamefont {Kaplan}, \citenamefont {Ning}, \citenamefont {Perdew},\ and\ \citenamefont {Sun}}]{2020_Furness_r2SCAN_functional}%
  \BibitemOpen
  \bibfield  {author} {\bibinfo {author} {\bibfnamefont {J.~W.}\ \bibnamefont {Furness}}, \bibinfo {author} {\bibfnamefont {A.~D.}\ \bibnamefont {Kaplan}}, \bibinfo {author} {\bibfnamefont {J.}~\bibnamefont {Ning}}, \bibinfo {author} {\bibfnamefont {J.~P.}\ \bibnamefont {Perdew}}, \ and\ \bibinfo {author} {\bibfnamefont {J.}~\bibnamefont {Sun}},\ }\href {\doibase 10.1021/acs.jpclett.0c02405} {\bibfield  {journal} {\bibinfo  {journal} {J. Phys. Chem. Lett.}\ }\textbf {\bibinfo {volume} {11}},\ \bibinfo {pages} {8208} (\bibinfo {year} {2020})}\BibitemShut {NoStop}%
\bibitem [{\citenamefont {Heyd}\ \emph {et~al.}(2003)\citenamefont {Heyd}, \citenamefont {Scuseria},\ and\ \citenamefont {Ernzerhof}}]{2003_HSE_functional}%
  \BibitemOpen
  \bibfield  {author} {\bibinfo {author} {\bibfnamefont {J.}~\bibnamefont {Heyd}}, \bibinfo {author} {\bibfnamefont {G.~E.}\ \bibnamefont {Scuseria}}, \ and\ \bibinfo {author} {\bibfnamefont {M.}~\bibnamefont {Ernzerhof}},\ }\href {\doibase 10.1063/1.1564060} {\bibfield  {journal} {\bibinfo  {journal} {J. Chem. Phys.}\ }\textbf {\bibinfo {volume} {118}},\ \bibinfo {pages} {8207} (\bibinfo {year} {2003})}\BibitemShut {NoStop}%
\bibitem [{\citenamefont {Becke}(1993)}]{1993_Becke_B3LYP_part1}%
  \BibitemOpen
  \bibfield  {author} {\bibinfo {author} {\bibfnamefont {A.~D.}\ \bibnamefont {Becke}},\ }\href {\doibase 10.1063/1.464913} {\bibfield  {journal} {\bibinfo  {journal} {J. Chem. Phys.}\ }\textbf {\bibinfo {volume} {98}},\ \bibinfo {pages} {5648} (\bibinfo {year} {1993})}\BibitemShut {NoStop}%
\bibitem [{\citenamefont {Stephens}\ \emph {et~al.}(1994)\citenamefont {Stephens}, \citenamefont {Devlin}, \citenamefont {Chabalowski},\ and\ \citenamefont {Frisch}}]{1994_Stephens_B3LYP_part2}%
  \BibitemOpen
  \bibfield  {author} {\bibinfo {author} {\bibfnamefont {P.~J.}\ \bibnamefont {Stephens}}, \bibinfo {author} {\bibfnamefont {F.~J.}\ \bibnamefont {Devlin}}, \bibinfo {author} {\bibfnamefont {C.~F.}\ \bibnamefont {Chabalowski}}, \ and\ \bibinfo {author} {\bibfnamefont {M.~J.}\ \bibnamefont {Frisch}},\ }\href {\doibase 10.1021/j100096a001} {\bibfield  {journal} {\bibinfo  {journal} {J. Phys. Chem.}\ }\textbf {\bibinfo {volume} {98}},\ \bibinfo {pages} {11623} (\bibinfo {year} {1994})}\BibitemShut {NoStop}%
\bibitem [{\citenamefont {Kresse}\ and\ \citenamefont {Furthm\"uller}(1996)}]{1996_Kresse_VASP_PRB}%
  \BibitemOpen
  \bibfield  {author} {\bibinfo {author} {\bibfnamefont {G.}~\bibnamefont {Kresse}}\ and\ \bibinfo {author} {\bibfnamefont {J.}~\bibnamefont {Furthm\"uller}},\ }\href {\doibase 10.1103/PhysRevB.54.11169} {\bibfield  {journal} {\bibinfo  {journal} {Phys. Rev. B}\ }\textbf {\bibinfo {volume} {54}},\ \bibinfo {pages} {11169} (\bibinfo {year} {1996})}\BibitemShut {NoStop}%
\bibitem [{\citenamefont {Bl\"ochl}(1994)}]{Blochl1994_PAW}%
  \BibitemOpen
  \bibfield  {author} {\bibinfo {author} {\bibfnamefont {P.~E.}\ \bibnamefont {Bl\"ochl}},\ }\href {\doibase 10.1103/PhysRevB.50.17953} {\bibfield  {journal} {\bibinfo  {journal} {Phys. Rev. B}\ }\textbf {\bibinfo {volume} {50}},\ \bibinfo {pages} {17953} (\bibinfo {year} {1994})}\BibitemShut {NoStop}%
\bibitem [{\citenamefont {Dudarev}\ \emph {et~al.}(1998)\citenamefont {Dudarev}, \citenamefont {Botton}, \citenamefont {Savrasov}, \citenamefont {Humphreys},\ and\ \citenamefont {Sutton}}]{1998_Dudarev_DFT+U_LDAUTYPE2}%
  \BibitemOpen
  \bibfield  {author} {\bibinfo {author} {\bibfnamefont {S.~L.}\ \bibnamefont {Dudarev}}, \bibinfo {author} {\bibfnamefont {G.~A.}\ \bibnamefont {Botton}}, \bibinfo {author} {\bibfnamefont {S.~Y.}\ \bibnamefont {Savrasov}}, \bibinfo {author} {\bibfnamefont {C.~J.}\ \bibnamefont {Humphreys}}, \ and\ \bibinfo {author} {\bibfnamefont {A.~P.}\ \bibnamefont {Sutton}},\ }\href {\doibase 10.1103/PhysRevB.57.1505} {\bibfield  {journal} {\bibinfo  {journal} {Phys. Rev. B}\ }\textbf {\bibinfo {volume} {57}},\ \bibinfo {pages} {1505} (\bibinfo {year} {1998})}\BibitemShut {NoStop}%
\bibitem [{\citenamefont {Ashkenazi}\ \emph {et~al.}(1989)\citenamefont {Ashkenazi}, \citenamefont {Pickett}, \citenamefont {Krakauer}, \citenamefont {Wang}, \citenamefont {Klein},\ and\ \citenamefont {Chubb}}]{1989_Ashkenazi_polyacetylene_Peierls}%
  \BibitemOpen
  \bibfield  {author} {\bibinfo {author} {\bibfnamefont {J.}~\bibnamefont {Ashkenazi}}, \bibinfo {author} {\bibfnamefont {W.~E.}\ \bibnamefont {Pickett}}, \bibinfo {author} {\bibfnamefont {H.}~\bibnamefont {Krakauer}}, \bibinfo {author} {\bibfnamefont {C.~S.}\ \bibnamefont {Wang}}, \bibinfo {author} {\bibfnamefont {B.~M.}\ \bibnamefont {Klein}}, \ and\ \bibinfo {author} {\bibfnamefont {S.~R.}\ \bibnamefont {Chubb}},\ }\href {\doibase 10.1103/PhysRevLett.62.2016} {\bibfield  {journal} {\bibinfo  {journal} {Phys. Rev. Lett.}\ }\textbf {\bibinfo {volume} {62}},\ \bibinfo {pages} {2016} (\bibinfo {year} {1989})}\BibitemShut {NoStop}%
\bibitem [{\citenamefont {Sun}\ \emph {et~al.}(2002)\citenamefont {Sun}, \citenamefont {Kürti}, \citenamefont {Kertesz},\ and\ \citenamefont {Baughman}}]{2002_Sun_polyacetylene_gating_DFT}%
  \BibitemOpen
  \bibfield  {author} {\bibinfo {author} {\bibfnamefont {G.}~\bibnamefont {Sun}}, \bibinfo {author} {\bibfnamefont {J.}~\bibnamefont {Kürti}}, \bibinfo {author} {\bibfnamefont {M.}~\bibnamefont {Kertesz}}, \ and\ \bibinfo {author} {\bibfnamefont {R.~H.}\ \bibnamefont {Baughman}},\ }\href {\doibase 10.1063/1.1509052} {\bibfield  {journal} {\bibinfo  {journal} {J. Chem. Phys.}\ }\textbf {\bibinfo {volume} {117}},\ \bibinfo {pages} {7691} (\bibinfo {year} {2002})}\BibitemShut {NoStop}%
\bibitem [{\citenamefont {Lechner}\ \emph {et~al.}(2016)\citenamefont {Lechner}, \citenamefont {Pannier}, \citenamefont {Baranek}, \citenamefont {Forero-Martinez},\ and\ \citenamefont {Vach}}]{2016_Lechner_carbon_DFT_benchmarking}%
  \BibitemOpen
  \bibfield  {author} {\bibinfo {author} {\bibfnamefont {C.}~\bibnamefont {Lechner}}, \bibinfo {author} {\bibfnamefont {B.}~\bibnamefont {Pannier}}, \bibinfo {author} {\bibfnamefont {P.}~\bibnamefont {Baranek}}, \bibinfo {author} {\bibfnamefont {N.~C.}\ \bibnamefont {Forero-Martinez}}, \ and\ \bibinfo {author} {\bibfnamefont {H.}~\bibnamefont {Vach}},\ }\href {\doibase 10.1021/acs.jpcc.5b10396} {\bibfield  {journal} {\bibinfo  {journal} {J. Phys. Chem. C}\ }\textbf {\bibinfo {volume} {120}},\ \bibinfo {pages} {5083} (\bibinfo {year} {2016})}\BibitemShut {NoStop}%
\bibitem [{\citenamefont {Fradkin}\ \emph {et~al.}(2015)\citenamefont {Fradkin}, \citenamefont {Kivelson},\ and\ \citenamefont {Tranquada}}]{2015_Tranquada_high-Tc_intertwined_orders}%
  \BibitemOpen
  \bibfield  {author} {\bibinfo {author} {\bibfnamefont {E.}~\bibnamefont {Fradkin}}, \bibinfo {author} {\bibfnamefont {S.~A.}\ \bibnamefont {Kivelson}}, \ and\ \bibinfo {author} {\bibfnamefont {J.~M.}\ \bibnamefont {Tranquada}},\ }\href {\doibase 10.1103/RevModPhys.87.457} {\bibfield  {journal} {\bibinfo  {journal} {Rev. Mod. Phys.}\ }\textbf {\bibinfo {volume} {87}},\ \bibinfo {pages} {457} (\bibinfo {year} {2015})}\BibitemShut {NoStop}%
\bibitem [{\citenamefont {Fernandes}\ \emph {et~al.}(2019)\citenamefont {Fernandes}, \citenamefont {Orth},\ and\ \citenamefont {Schmalian}}]{2019_Schmalian_vestigal_order_in_QM}%
  \BibitemOpen
  \bibfield  {author} {\bibinfo {author} {\bibfnamefont {R.~M.}\ \bibnamefont {Fernandes}}, \bibinfo {author} {\bibfnamefont {P.~P.}\ \bibnamefont {Orth}}, \ and\ \bibinfo {author} {\bibfnamefont {J.}~\bibnamefont {Schmalian}},\ }\href {\doibase 10.1146/annurev-conmatphys-031218-013200} {\bibfield  {journal} {\bibinfo  {journal} {Annu. Rev. Condens. Matter Phys.}\ }\textbf {\bibinfo {volume} {10}},\ \bibinfo {pages} {133} (\bibinfo {year} {2019})}\BibitemShut {NoStop}%
\bibitem [{\citenamefont {Kivelson}\ and\ \citenamefont {Emery}(1996)}]{1996_Kivelson_topological_doping_polyacetylene}%
  \BibitemOpen
  \bibfield  {author} {\bibinfo {author} {\bibfnamefont {S.~A.}\ \bibnamefont {Kivelson}}\ and\ \bibinfo {author} {\bibfnamefont {V.~J.}\ \bibnamefont {Emery}},\ }\href {\doibase 10.1016/S0379-6779(96)03696-X} {\bibfield  {journal} {\bibinfo  {journal} {Synth. Met.}\ }\textbf {\bibinfo {volume} {80}},\ \bibinfo {pages} {151} (\bibinfo {year} {1996})}\BibitemShut {NoStop}%
\bibitem [{\citenamefont {Mej\'{\i}a-Rodr\'{\i}guez}\ and\ \citenamefont {Trickey}(2019)}]{2019_Trickey_SCAN_TM_overmagnetization}%
  \BibitemOpen
  \bibfield  {author} {\bibinfo {author} {\bibfnamefont {D.}~\bibnamefont {Mej\'{\i}a-Rodr\'{\i}guez}}\ and\ \bibinfo {author} {\bibfnamefont {S.~B.}\ \bibnamefont {Trickey}},\ }\href {\doibase 10.1103/PhysRevB.100.041113} {\bibfield  {journal} {\bibinfo  {journal} {Phys. Rev. B}\ }\textbf {\bibinfo {volume} {100}},\ \bibinfo {pages} {041113} (\bibinfo {year} {2019})}\BibitemShut {NoStop}%
\bibitem [{\citenamefont {Tran}\ \emph {et~al.}(2020)\citenamefont {Tran}, \citenamefont {Baudesson}, \citenamefont {Carrete}, \citenamefont {Madsen}, \citenamefont {Blaha}, \citenamefont {Schwarz},\ and\ \citenamefont {Singh}}]{2020_Singh_SCAN_magnetism_shortcomings}%
  \BibitemOpen
  \bibfield  {author} {\bibinfo {author} {\bibfnamefont {F.}~\bibnamefont {Tran}}, \bibinfo {author} {\bibfnamefont {G.}~\bibnamefont {Baudesson}}, \bibinfo {author} {\bibfnamefont {J.}~\bibnamefont {Carrete}}, \bibinfo {author} {\bibfnamefont {G.~K.~H.}\ \bibnamefont {Madsen}}, \bibinfo {author} {\bibfnamefont {P.}~\bibnamefont {Blaha}}, \bibinfo {author} {\bibfnamefont {K.}~\bibnamefont {Schwarz}}, \ and\ \bibinfo {author} {\bibfnamefont {D.~J.}\ \bibnamefont {Singh}},\ }\href {\doibase 10.1103/PhysRevB.102.024407} {\bibfield  {journal} {\bibinfo  {journal} {Phys. Rev. B}\ }\textbf {\bibinfo {volume} {102}},\ \bibinfo {pages} {024407} (\bibinfo {year} {2020})}\BibitemShut {NoStop}%
\bibitem [{\citenamefont {Cannuccia}\ and\ \citenamefont {Marini}(2011)}]{2011_Cannuccia_polyacetylene_zero_point_motion}%
  \BibitemOpen
  \bibfield  {author} {\bibinfo {author} {\bibfnamefont {E.}~\bibnamefont {Cannuccia}}\ and\ \bibinfo {author} {\bibfnamefont {A.}~\bibnamefont {Marini}},\ }\href {\doibase 10.1103/PhysRevLett.107.255501} {\bibfield  {journal} {\bibinfo  {journal} {Phys. Rev. Lett.}\ }\textbf {\bibinfo {volume} {107}},\ \bibinfo {pages} {255501} (\bibinfo {year} {2011})}\BibitemShut {NoStop}%
\end{thebibliography}
\end{document}